\documentclass{article}

\usepackage{ITW06_PaperLaTeXStyle}

\usepackage{IEEEtrantools}
\usepackage{graphicx}
\usepackage{subfigure}

\usepackage{bm}
\usepackage[noadjust]{cite}

\newtheorem{theorem}{Theorem}
\newtheorem{lemma}{Lemma}
\newtheorem{proposition}{Proposition}

\newcommand{\eqref}[1]{\textnormal{(\ref{#1})}}
\newcommand{\relvar}[2]{\buildrel {#2} \over {#1}}
\newcommand{\eqvar}[1]{\relvar{=}{#1}}

\newcommand{\levar}[1]{\relvar{\le}{#1}}

\newcommand{\eqdef}{\eqvar{\Delta}}

\def\QEDopen{{\setlength{\fboxsep}{0pt}\setlength{\fboxrule}{0.2pt}\fbox{\rule[0pt]{0pt}{1.3ex}\rule[0pt]{1.3ex}{0pt}}}}
\newcommand{\QED}{\QEDopen}
\newenvironment{proofof}[2][Proof of]{\noindent\hspace{2em}{\itshape #1 {#2}: }}{\hspace*{\fill}~\QED\par\endtrivlist\unskip}

\begin{document}
\headheight = 10mm
\headsep = 6mm

\itwtitle{On the Performance of Lossless Joint Source-Channel Coding Based on Linear Codes}


\itwauthorswithsameaddress{Shengtian Yang, Peiliang Qiu\footnotemark[1]}
{Department of Information Science \& Electronic Engineering\\
Zhejiang University\\
Hangzhou, Zhejiang 310027, China\\
{\tt \{yangshengtian, qiupl\}@zju.edu.cn}}

\markboth{}{To appear in Proc. 2006 IEEE Information Theory Workshop, October 22-26, 2006, Chengdu, China.}
\pagestyle{myheadings}

\itwmaketitle
\thispagestyle{myheadings}

\footnotetext[1]{This work was supported in part by the Natural Science Foundation of China under Grant NSFC-60472079 and by the Chinese Specialized Research Fund for the Doctoral Program of Higher Education under Grant 2004-0335099.}

\begin{itwabstract}
A general lossless joint source-channel coding scheme based on linear codes is proposed and then analyzed in this paper. It is shown that a linear code with good joint spectrum can be used to establish limit-approaching joint source-channel coding schemes for arbitrary sources and channels, where the joint spectrum of the code is a generalization of the input-output weight distribution.
\end{itwabstract}

\begin{itwpaper}

\itwsection{Introduction}

In a traditional communication system, source and channel coding are treated independently. This is because Shannon in 1948 showed that separate source and channel coding incurs no loss of optimality provided that the coding length goes to infinity, which is now called the separation theorem. While this separation is well motivated for the point-to-point case, it can entail significant performance losses in more general scenarios, for example, the transmission of correlated sources over multiple access channels \cite{JSCC:Cover198011}. Moreover, even for the point-to-point case, it is recently reported by Zhong et al \cite{JSCC:Zhong200604} that the joint source-channel coding usually works more efficiently (in terms of the error exponent) than does the separate coding. Therefore, in many applications, it is expected to adopt a joint source-channel coding scheme.

However, for arbitrary sources and channels, how to construct a limit-approaching joint source-channel coding scheme? To answer this question, we propose and analyze a general lossless joint source-channel coding scheme based on linear codes in this paper.

\itwsection{Linear Codes, Types and Spectrums}

Before presenting our coding scheme, we first need to introduce some new concepts, definitions and notations related to linear codes. Let $\mathcal{X}$ and $\mathcal{Y}$ be two finite additive groups, then a \emph{linear code} can be defined by a homomorphism $f: \mathcal{X}^n \to \mathcal{Y}^m$, i.e., a map satisfying
$$
f(\bm{x}_1 + \bm{x}_2) = f(\bm{x}_1) + f(\bm{x}_2), \quad \forall \bm{x}_1, \bm{x}_2 \in \mathcal{X}^n
$$
where $\mathcal{X}^n$ and $\mathcal{Y}^m$ denote the direct product of $n$ groups $\mathcal{X}$ and $m$ groups $\mathcal{Y}$, respectively. Note that any permutation $\sigma_n$ on $n$ letters can be regarded as an automorphism on $\mathcal{X}^n$, and we denote by $\Sigma_n$ an independent uniform random permutation on $n$ letters.

Next, we introduce the concept of types in the methods of types \cite{JSCC:Csiszar198100}. The \emph{type} of a sequence $\bm{x}$ in $\mathcal{X}^n$ is the empirical distribution $P_{\bm{x}}$ on $\mathcal{X}$ defined by
$$
P_{\bm{x}}(a) \eqdef \frac{N(a|\bm{x})}{|\bm{x}|}, \quad \forall a \in \mathcal{X}
$$
where $N(a|\bm{x})$ denotes the number of occurrences of $a$ in $\bm{x}$ and $|\bm{x}|$ denotes the length of $\bm{x}$. For any distribution $P$ on $\mathcal{X}$, the set of sequences of type $P$ in $\mathcal{X}^n$ is denoted by $\mathcal{T}_P^n(\mathcal{X})$. A distribution $P$ on $\mathcal{X}$ is called a type of sequences in $\mathcal{X}^n$ if $\mathcal{T}_P^n(\mathcal{X}) \ne \emptyset$. We denote by $\mathcal{P}(\mathcal{X})$ the set of all distributions on $\mathcal{X}$, and denote by $\mathcal{P}_n(\mathcal{X})$ the set of all possible types of sequences in $\mathcal{X}^n$.

Now we start to define the spectrum. The \emph{spectrum} of a set $A \subseteq \mathcal{X}^n$ is the empirical distribution $S_{\mathcal{X}}(A)$ on $\mathcal{P}(\mathcal{X})$ defined by
$$
S_{\mathcal{X}}(A)(P) \eqdef \frac{|\{\bm{x} \in A | P_{\bm{x}} = P\}|}{|A|}, \quad \forall P \in \mathcal{P}(\mathcal{X}).
$$
Similarly, the \emph{joint spectrum} of a set $B \subseteq \mathcal{X}^n \times \mathcal{Y}^m$ is the empirical distribution $S_{\mathcal{X}\mathcal{Y}}(B)$ on $\mathcal{P}(\mathcal{X}) \times \mathcal{P}(\mathcal{Y})$ defined by
$$
S_{\mathcal{X}\mathcal{Y}}(B)(P, Q) \eqdef \frac{|\{(\bm{x}, \bm{y}) \in B | P_{\bm{x}} = P, P_{\bm{y}} = Q\}|}{|B|},
$$
for all $P \in \mathcal{P}(\mathcal{X}), Q \in \mathcal{P}(\mathcal{Y})$. Furthermore, the \emph{marginal spectrums} and the \emph{conditional spectrums} of $B$ can be defined as the marginal distributions and the conditional distributions of $S_{\mathcal{X}\mathcal{Y}}(B)$, respectively, that is,
$$
S_{\mathcal{X}}(B)(P) \eqdef \sum_{Q \in \mathcal{P}(\mathcal{Y})} S_{\mathcal{X}\mathcal{Y}}(B)(P, Q),
$$
$$
S_{\mathcal{Y}}(B)(Q) \eqdef \sum_{P \in \mathcal{P}(\mathcal{X})} S_{\mathcal{X}\mathcal{Y}}(B)(P, Q),
$$
$$
S_{\mathcal{Y}|\mathcal{X}}(B)(Q|P) \eqdef \frac{S_{\mathcal{X}\mathcal{Y}}(B)(P, Q)}{S_{\mathcal{X}}(B)(P)},
$$
$$
S_{\mathcal{X}|\mathcal{Y}}(B)(P|Q) \eqdef \frac{S_{\mathcal{X}\mathcal{Y}}(B)(P, Q)}{S_{\mathcal{Y}}(B)(Q)}.
$$
Note that the conditional spectrum $S_{\mathcal{Y}|\mathcal{X}}(B)(Q|P)$ (or $S_{\mathcal{X}|\mathcal{Y}}(B)(P|Q)$) is well defined only for those $P$ (or $Q$) satisfying $S_{\mathcal{X}}(B)(P) \ne 0$ (or $S_{\mathcal{Y}}(B)(Q) \ne 0$).

Then naturally, for a given function $f: \mathcal{X}^n \to \mathcal{Y}^m$, we may define its \emph{joint spectrum} $S_{\mathcal{X}\mathcal{Y}}(f)$, \emph{forward conditional spectrum} $S_{\mathcal{Y}|\mathcal{X}}(f)$ and \emph{image spectrum} $S_{\mathcal{Y}}(f)$ as $S_{\mathcal{X}\mathcal{Y}}(\mathrm{rl}(f))$, $S_{\mathcal{Y}|\mathcal{X}}(\mathrm{rl}(f))$ and $S_{\mathcal{Y}}(\mathrm{rl}(f))$, respectively, where $\mathrm{rl}(f)$ is the \emph{relation} defined by $\{(\bm{x}, f(\bm{x})) | \bm{x} \in \mathcal{X}^n\}$. Careful readers must have noticed that the image spectrum $S_{\mathcal{Y}}(f)$ and the joint spectrum $S_{\mathcal{X}\mathcal{Y}}(f)$ are virtually the generalized and normalized versions of the spectrum (e.g., \cite{JSCC:Bennatan200403}) and the input-output weight distribution (e.g., \cite{JSCC:Divsalar199809}), respectively.

From the above definitions, we can easily obtain the following properties.

\begin{proposition}
$$
S_{\mathcal{X}}(\mathcal{X}^n)(P) = \frac{{n \choose nP}}{|\mathcal{X}|^n}, \quad S_{\mathcal{Y}}(\mathcal{Y}^m)(Q) = \frac{{m \choose mQ}}{|\mathcal{Y}|^n},
$$
$$
S_{\mathcal{X}\mathcal{Y}}(\mathcal{X}^n \times \mathcal{Y}^m)(P, Q) = S_{\mathcal{X}}(\mathcal{X}^n)(P) \cdot S_{\mathcal{Y}}(\mathcal{Y}^m)(Q)
$$
for any $P \in \mathcal{P}_n(\mathcal{X})$, $Q \in \mathcal{P}_m(\mathcal{Y})$, where
$$
{n \choose nP} \eqdef \frac{n!}{\prod_{a \in \mathcal{X}} (nP(a))!}, \quad {m \choose mQ} \eqdef \frac{m!}{\prod_{b \in \mathcal{Y}} (mQ(b))!}.
$$
\end{proposition}

\begin{proposition}
For a given function $f: \mathcal{X}^n \to \mathcal{Y}^m$,
$$
S_{\mathcal{X}\mathcal{Y}}(\sigma_m \circ f \circ \sigma_n) = S_{\mathcal{X}\mathcal{Y}}(f).
$$
for any permutations $\sigma_n$ and $\sigma_m$
\end{proposition}

\begin{proposition}
For a given random function $F: \mathcal{X}^n \to \mathcal{Y}^m$, we have
$$
\Pr\{\tilde{F}(\bm{x}) = \bm{y}\} = |\mathcal{Y}|^{-m} \alpha(F)(P_{\bm{x}}, P_{\bm{y}})
$$
for any $\bm{x} \in \mathcal{X}^n$, $\bm{y} \in \mathcal{Y}^m$, where
\begin{equation}\label{eq:RandomizedF1}
\tilde{F} \eqdef \Sigma_m \circ F \circ \Sigma_n
\end{equation}
and
$$
\alpha(F)(P, Q) \eqdef \frac{E[S_{\mathcal{X}\mathcal{Y}}(F)(P, Q)]}{S_{\mathcal{X}\mathcal{Y}}(\mathcal{X}^n \times \mathcal{Y}^m)(P, Q)}.
$$
\end{proposition}

\begin{proposition}\label{pr:GoodLinearCodes}
If both $\mathcal{X}$ and $\mathcal{Y}$ are Galois field $\mathrm{GF}(q)$, then there exists a random linear code $F: \mathcal{X}^n \to \mathcal{Y}^m$ such that
$$
\Pr\{F(\bm{x}) = \bm{y}\} = |\mathcal{Y}|^{-m}
$$
for any $\bm{x} \in \mathcal{X}^n$ $(\bm{x} \ne 0^n)$, $\bm{y} \in \mathcal{Y}^m$, or equivalently
$$
\alpha(F)(P, Q) = 1
$$
for all $P \in \mathcal{P}_n(\mathcal{X})$ $(P \ne P_{0^n})$ and all $Q \in \mathcal{P}_m(\mathcal{Y})$.
\end{proposition}

\begin{proposition}\label{pr:IndependenceLC1}
For a given random linear code $F: \mathcal{X}^n \to \mathcal{Y}^m$, we have
\begin{IEEEeqnarray*}{l}
\Pr\{\hat{F}(\bm{x}_1) = \bm{y}_1\} = |\mathcal{Y}|^{-m}, \\
\Pr\{\hat{F}(\bm{x}_2) = \bm{y}_2 | \hat{F}(\bm{x}_1) = \bm{y}_1\} \\
\quad = |\mathcal{Y}|^{-m} \alpha(F)(P_{\bm{x}_2-\bm{x}_1}, P_{\bm{y}_2-\bm{y}_1}),
\end{IEEEeqnarray*}
for any unequal $\bm{x}_1, \bm{x}_2 \in \mathcal{X}^n$ and any $\bm{y}_1, \bm{y}_2 \in \mathcal{Y}^m$, where
\begin{equation}\label{eq:RandomizedF2}
\hat{F}(\bm{x}) \eqdef \tilde{F}(\bm{x}) + \bar{Y}^m
\end{equation}
where $\bar{Y}^m$ denotes a uniform random vector on $\mathcal{Y}^m$.
\end{proposition}

\itwsection{Lossless Joint Source-Channel Coding Based on Linear Codes}

In this section, we will present a lossless joint source-channel coding scheme based on linear codes for arbitrary sources and channels. To analyze the performance of the scheme, we use the information-spectrum methods \cite[pp. 247-268]{JSCC:Han200300}. In the methods of information-spectrum, a \emph{general source} is defined as an infinite sequence
$$
\bm{X} = \{X^n = (X_1^{(n)}, X_2^{(n)}, \cdots, X_n^{(n)})\}_{n=1}^{\infty}
$$
of $n$-dimensional random variables $X^n$ where each component random variable $X_i^{(n)}$ ($1 \le i \le n$) takes values in the alphabet $\mathcal{X}$, and a \emph{general channel} is defined as an infinite sequence $\bm{W} = \{W^n\}_{n=1}^{\infty}$ of conditional probability distribution $W^n = W^n(\cdot|\cdot)$ satisfying
$$
\sum_{\bm{y} \in \mathcal{Y}^n} W^n(\bm{y}|\bm{x}) = 1, \quad \forall \bm{x} \in \mathcal{X}^n
$$
for each $n = 1, 2, \cdots$. Then by the similar notations and definitions in \cite[pp. 247-268]{JSCC:Han200300}, a general joint source-channel coding problem may be described as follows.

First, let $\mathcal{V}$ be a source alphabet (a finite additive group). Denote by $\mathcal{X}$ and $\mathcal{Y}$ an input alphabet and an output alphabet of a channel, respectively ($\mathcal{X}$ and $\mathcal{Y}$ can be arbitrary sets). Suppose that an arbitrary general source $\bm{V} = \{V^n\}_{n=1}^\infty$ and an arbitrary general channel $\bm{W} = \{W^n\}_{n=1}^\infty$ are given. We define an \emph{encoder} $\varphi_n: \mathcal{V}^n \to \mathcal{X}^m$ and a \emph{decoder} $\psi_n: \mathcal{Y}^m \to \mathcal{V}^n$ as arbitrary mappings. Setting $X^m = \varphi_n(V^n)$ and denoting by $Y^m$ the output from the channel $W^m$ with $X^m$ as the input, the lossless joint source-channel coding system is virtually a Markov chain
$$
V^n \relvar{\longrightarrow}{\varphi_n} X^m \relvar{\longrightarrow}{W^m} Y^m \longrightarrow \psi_n(Y^m).
$$
We define the \emph{error probability} $\epsilon_n$ by
\begin{IEEEeqnarray*}{rCl}
\epsilon_n &\eqdef &\Pr\{V^n \ne \psi_n(Y^m)\} \\
&= &\sum_{\bm{v} \in \mathcal{V}^n} P_{V^n}(\bm{v}) W^m([\psi_n^{-1}(\bm{v})]^c | \varphi_n(\bm{v})),
\end{IEEEeqnarray*}
that is, $\epsilon_n$ is defined as the average error probability with respect to the probability distribution $P_{V^n}$ of the source. For simplicity, we call a pair $(\varphi_n, \psi_n)$ of an encoder and a decoder with the error probability $\epsilon_n$ an $(n, m, \epsilon_n)$-code. Then for a general source $\bm{V}$ and a general channel $\bm{W}$, we define a source $\bm{V}$ being $(R, \epsilon)$-transmissible over channel $\bm{W}$ if there exists an $(n, m_n, \epsilon_n)$-code satisfying
$$
\limsup_{n \to \infty} \frac{n}{m_n} \le R, \quad \limsup_{n \to \infty} \epsilon_n \le \epsilon.
$$

Now let us consider our coding scheme based on linear codes, which is depicted in Figure \ref{fig:Scheme1}.
\begin{figure*}[htb]
\centering
\includegraphics{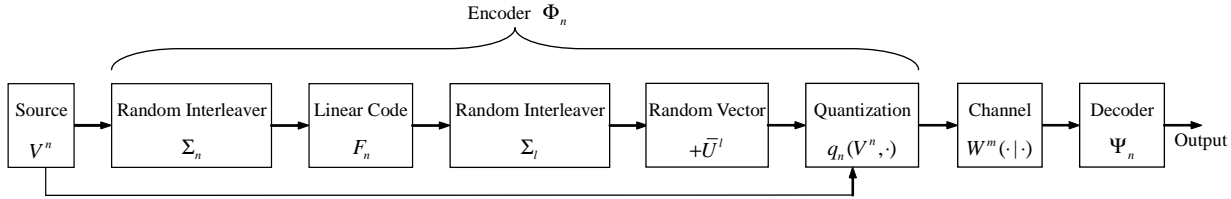}
\caption{The proposed lossless joint source-channel coding scheme based on linear codes}
\label{fig:Scheme1}
\end{figure*}
According to the scheme, the (random) encoder $\Phi_n$ is defined by
\begin{equation}\label{eq:EncoderDefinition}
\Phi_n(\bm{v}) = q_n(\bm{v}, \hat{F}_n(\bm{v})) = q_n(\bm{v}, \Sigma_l(F_n(\Sigma_n(\bm{v}))) + \bar{U}^l),
\end{equation}
where $F_n: \mathcal{V}^n \to \mathcal{U}^l$ is a random linear code ($\mathcal{U}$ is a finite additive group) and $\hat{F}_n$ is defined by \eqref{eq:RandomizedF1} and \eqref{eq:RandomizedF2}, and the quantization $q_n$ is a map from $\mathcal{V}^n \times \mathcal{U}^l$ to $\mathcal{X}^m$.
For comparison, the channel coding scheme \cite{JSCC:Bennatan200403} and the lossless source coding scheme \cite{JSCC:Yang200503, JSCC:Muramatsu200510} based on linear codes are shown in Figure \ref{fig:Scheme2}.
\begin{figure*}[htb]
\centering
\subfigure[Channel Coding]{\includegraphics{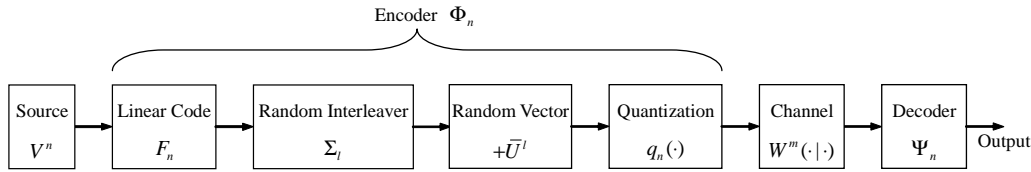}} 
\subfigure[Lossless Source Coding]{\includegraphics[width=4.2in]{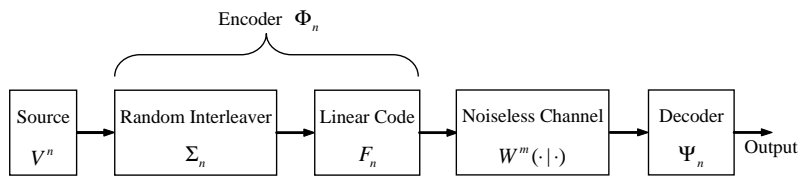}}
\caption{The channel coding scheme and the lossless source coding scheme based on linear codes}
\label{fig:Scheme2}
\end{figure*}
Note that our scheme is in fact a combination of the two schemes except that the quantization $q_n$ is now modified to be correlated with the source output $V^n$.

Next, let us investigate the $(R, \epsilon)$-transmissible condition of our scheme based on the random linear code $F_n$. To this end, we need Lemma \ref{le:IndependenceLC2} and Lemma \ref{le:FeinsteinBLC}.

\begin{lemma}\label{le:IndependenceLC2}
For a given random linear code $F_n: \mathcal{V}^n \to \mathcal{U}^l$ and a given quantization $q_n: \mathcal{V}^n \times \mathcal{U}^l \to \mathcal{X}^m$, we have
\begin{IEEEeqnarray}{l}
\Pr\{\Phi_n(\bm{v}_1) = \bm{x}_1\} = P_{X^m|V^n}(\bm{x}_1|\bm{v}_1), \\
\Pr\{\Phi_n(\bm{v}_2) = \bm{x}_2 | \Phi_n(\bm{v}_1) = \bm{x}_1\} \IEEEnonumber \\
= \beta(F_n, q_n)(\bm{v}_1, \bm{v}_2, \bm{x}_1, \bm{x}_2) P_{X^m|V^n}(\bm{x}_2|\bm{v}_2) \\
\le \beta'(F_n, q_n)(\bm{v}_2, \bm{x}_2) P_{X^m|V^n}(\bm{x}_2|\bm{v}_2)
\end{IEEEeqnarray}
for any $\bm{v}_1, \bm{v}_2 \in \mathcal{V}^n$ $(\bm{v}_1 \ne \bm{v}_2)$ and any $\bm{x}_1, \bm{x}_2 \in \mathcal{X}^m$, where $\Phi_n$ is defined by \eqref{eq:EncoderDefinition}, and
\begin{equation}\label{eq:RequirementOfQ}
P_{X^m|V^n}(\bm{x}|\bm{v}) = \frac{|q_n^{-1}(\bm{v}, \bm{x})|}{|\mathcal{U}|^l},
\end{equation}
\begin{equation}
q_n^{-1}(\bm{v}, \bm{x}) \eqdef \{\bm{u} \in \mathcal{U}^l | q_n(\bm{v}, \bm{u}) = \bm{x}\},
\end{equation}
\begin{IEEEeqnarray}{l}
\beta(F_n, q_n)(\bm{v}_1, \bm{v}_2, \bm{x}_1, \bm{x}_2) \eqdef \IEEEnonumber \\
\quad \sum_{\bm{u}_1 \in q_n^{-1}(\bm{v}_1, \bm{x}_1) \atop \bm{u}_2 \in q_n^{-1}(\bm{v}_2, \bm{x}_2)} \frac{\alpha(F_n)(P_{\bm{v}_2-\bm{v}_1}, P_{\bm{u}_2-\bm{u}_1})}{|q_n^{-1}(\bm{v}_1, \bm{x}_1)||q_n^{-1}(\bm{v}_2, \bm{x}_2)|},
\end{IEEEeqnarray}
\begin{IEEEeqnarray}{l}
\beta'(F_n, q_n)(\bm{v}_2, \bm{x}_2) \eqdef \IEEEnonumber \\
\max_{P \in \mathcal{P}_n(\mathcal{V}) \backslash \{P_{0^n}\} \atop \bm{u}_1 \in \mathcal{U}^l} \sum_{\bm{u}_2 \in q_n^{-1}(\bm{v}_2, \bm{x}_2)}  \frac{\alpha(F_n)(P, P_{\bm{u}_2-\bm{u}_1})}{|q_n^{-1}(\bm{v}_2, \bm{x}_2)|}.\label{eq:DefinitionOfBeta2}
\end{IEEEeqnarray}
\end{lemma}

Lemma \ref{le:IndependenceLC2} is an easy consequence of Proposition \ref{pr:IndependenceLC1} and hence its proof is omitted here, but it does play an important role for coding schemes based on linear codes. Though in most proofs of the theorems in information theory it is required to generate a sequence of independent variables, we only need pairwise independence in the proof of lossless joint source-channel (or channel) coding. Therefore, \emph{the art of lossless joint source-channel coding is how to generate a sequence of mutually independent variables subject to a conditional probability}, and Lemma \ref{le:IndependenceLC2} provides a feasible method based on linear codes for generating such sequences.

\begin{lemma}\label{le:FeinsteinBLC}
For a given random linear code $F_n: \mathcal{V}^n \to \mathcal{U}^l$ and a given quantization $q_n: \mathcal{V}^n \times \mathcal{U}^l \to \mathcal{X}^m$, the average error probability $\epsilon_n$ of the system (Figure \ref{fig:Scheme1}) based on $F_n$ and $q_n$ with optimal decoders satisfies
\begin{IEEEeqnarray}{rCl}
\epsilon_n &\le &\Pr\biggl\{\frac{1}{n} \ln \frac{W^m(Y^m|X^m)}{P_{Y^m}(Y^m)} \le \frac{1}{n} \ln \frac{1}{P_{V^n}(V^n)} \IEEEnonumber \\
& &+\: \frac{1}{n} \ln \beta'(F_n, q_n)(V^n, X^n) + \gamma \biggr\} + e^{-n\gamma}
\end{IEEEeqnarray}
where $X^m$ denotes the channel input generated randomly subject to the conditional probability distribution $P_{X^m|V^n}$ defined by \eqref{eq:RequirementOfQ} and $Y^m$ denotes the output from the channel $W^m$ corresponding to $X^m$, and $\beta'(F_n, q_n)$ is defined by \eqref{eq:DefinitionOfBeta2}.
\end{lemma}

As a ``linear code'' version of Lemma 3.8.1 in \cite{JSCC:Han200300}, Lemma \ref{le:FeinsteinBLC} is the most important result on the performance of our coding scheme based on linear codes. Its proof is presented as follows.

\begin{proofof}{Lemma \ref{le:FeinsteinBLC}}
By \eqref{eq:EncoderDefinition}, we have defined a random encoder $\Phi_n: \mathcal{V}^n \to \mathcal{X}^m$ based on the random linear code $F_n: \mathcal{V}^n \to \mathcal{U}^l$ and the quantization $q_n$, then it follows from Lemma \ref{le:IndependenceLC2} that
\begin{IEEEeqnarray}{l}
\Pr\{\Phi_n(\bm{v}) = \bm{x}\} = P_{X^m|V^n}(\bm{x}|\bm{v}), \label{eq:C1ofLemmaFeinsteinBLC} \\
\Pr\{\Phi_n(\bm{v}') = \bm{x}' | \Phi_n(\bm{v}) = \bm{x}\} \IEEEnonumber \\
\quad \le \beta'(F_n, q_n)(\bm{v}', \bm{x}') P_{X^m|V^n}(\bm{x}'|\bm{v}') \label{eq:C2ofLemmaFeinsteinBLC}
\end{IEEEeqnarray}
for any $\bm{v}, \bm{v}' \in \mathcal{V}^n$ $(\bm{v} \ne \bm{v}')$ and any $\bm{x}, \bm{x}' \in \mathcal{X}^m$.

In order to define a decoder $\psi_n: \mathcal{Y}^m \to \mathcal{V}^n$, we set
\begin{IEEEeqnarray}{rCl}
S_n &= &\biggl\{ (\bm{v}, \bm{x}, \bm{y}) \in \mathcal{V}^n \!\times\! \mathcal{X}^m \!\times\! \mathcal{Y}^m \bigg| \frac{1}{n} \ln \frac{W^m(\bm{y}|\bm{x})}{P_{Y^m}(\bm{y})} > \IEEEnonumber \\
& &\frac{1}{n} \ln \frac{1}{P_{V^n}(\bm{v})} + \frac{1}{n} \ln \beta'(F_n, q_n)(\bm{v}, \bm{x}) + \gamma \biggr\}, \label{eq:GoodSetForDecoding}
\end{IEEEeqnarray}
\begin{equation}\label{eq:GoodSet2ForDecoding}
S_n(\bm{v}) = \{(\bm{x}, \bm{y}) \in \mathcal{X}^m \!\times\! \mathcal{Y}^m | (\bm{v}, \bm{x}, \bm{y}) \in S_n\}.
\end{equation}
Suppose that a channel output $\bm{y} \in \mathcal{Y}^m$ is received, we define the decoder by $\bm{v} = \psi_n(\bm{y})$ if there exits a unique $\bm{v} \in \mathcal{V}^n$ satisfying $(\varphi_n(\bm{v}), \bm{y}) \in S_n(\bm{v})$. If there exists no such $\bm{v}$ or exist more than one such $\bm{v}$, we define $\psi_n(\bm{y}) = \bm{v}_0$, an arbitrary element in $\mathcal{V}^n$. Then for each sample encoder $\varphi_n$ generated by $\Phi_n$, there is a well defined decoder $\psi_n$, and we denote by $\Psi_n$ the whole random ensemble of the decoders with respect to the random encoder $\Phi_n$. The error probability $\epsilon_n$ with respect to the pair $(\Phi_n, \Psi_n)$ of the random encoder and decoder is then given by
\begin{equation}\label{eq:AverageErrorFormula}
\epsilon_n = \sum_{\bm{v} \in \mathcal{V}^n} P_{V^n}(\bm{v}) \epsilon_n(\bm{v}),
\end{equation}
where $\epsilon_n(\bm{v})$ denotes the error probability of a source output $\bm{v} \in \mathcal{V}^n$, and it can be bounded above in the following way:
\begin{IEEEeqnarray}{rCl}
\epsilon_n(\bm{v}) &\le &\Pr\{(\Phi_n(\bm{v}), Y^m) \not \in S_n(\bm{v})\} \IEEEnonumber \\
& &+\: \Pr\biggl\{\bigcup_{\bm{v}': \bm{v}' \ne \bm{v}} \{(\Phi_n(\bm{v}'), Y^m) \in S_n(\bm{v}')\} \biggr\} \IEEEnonumber \\
&\le &\Pr\{(\Phi_n(\bm{v}), Y^m) \not \in S_n(\bm{v})\} \IEEEnonumber \\
& &+\: \sum_{\bm{v}': \bm{v}' \ne \bm{v}} \Pr\{(\Phi_n(\bm{v}'), Y^m) \in S_n(\bm{v}')\}, \label{eq:UBofEpsilonV}
\end{IEEEeqnarray}
where $Y^m$ denotes the channel output corresponding to the input $\Phi_n(\bm{v})$. Since the first term on the right-hand side of \eqref{eq:UBofEpsilonV} can be written as
\begin{IEEEeqnarray*}{rCl}
A_n(\bm{v}) &\eqdef &\Pr\{(\Phi_n(\bm{v}), Y^m) \not \in S_n(\bm{v})\} \\
&= &\sum_{(\bm{x}, \bm{y}) \not \in S_n(\bm{v})} \Pr\{\Phi_n(\bm{v}) = \bm{x}\} W^m(\bm{y}|\bm{x}) \\
&\eqvar{(a)} &\sum_{(\bm{x}, \bm{y}) \not \in S_n(\bm{v})} P_{X^m|V^n}(\bm{x}|\bm{v}) W^m(\bm{y}|\bm{x}) \\
&= &\sum_{(\bm{x}, \bm{y}) \not \in S_n(\bm{v})} P_{X^mY^m|V^n}(\bm{x},\bm{y}|\bm{v}),
\end{IEEEeqnarray*}
where (a) follows from \eqref{eq:C1ofLemmaFeinsteinBLC}, it follows that
\begin{IEEEeqnarray}{Cl}
&\sum_{\bm{v} \in \mathcal{V}^n} P_{V^n}(\bm{v}) A_n(\bm{v}) \IEEEnonumber \\
= &\sum_{\bm{v} \in \mathcal{V}^n} P_{V^n}(\bm{v}) \sum_{(\bm{x}, \bm{y}) \not \in S_n(\bm{v})} P_{X^mY^m|V^n}(\bm{x},\bm{y}|\bm{v}) \IEEEnonumber \\
= &\sum_{(\bm{v}, \bm{x}, \bm{y}) \not \in S_n} P_{V^nX^mY^m}(\bm{v}, \bm{x},\bm{y}) \IEEEnonumber \\
= &\Pr\{(V^n, X^m, Y^m) \not \in S_n\}.\label{eq:UBofAverageAn}
\end{IEEEeqnarray}
On the other hand, the second term on the right-hand side of \eqref{eq:UBofEpsilonV} can be written as
\begin{IEEEeqnarray*}{Cl}
&B_n(\bm{v}) \\
\eqdef &\sum_{\bm{v}': \bm{v}' \ne \bm{v}} \Pr\{(\Phi_n(\bm{v}'), Y^m) \in S_n(\bm{v}')\} \\
= &\sum_{\bm{v}': \bm{v}' \ne \bm{v}} \sum_{(\bm{x}, \bm{y}) \in \mathcal{X}^m \times \mathcal{Y}^m} \Pr\{\Phi_n(\bm{v}) = \bm{x}\} W^m(\bm{y}|\bm{x}) \\
&\sum_{\bm{x}' \in \mathcal{X}^m}\! \Pr\{\Phi_n(\bm{v}') = \bm{x}' | \Phi_n(\bm{v}) = \bm{x}\} 1\{(\bm{x}', \bm{y}) \in S_n(\bm{v}')\} \\
\levar{(a)} &\sum_{\bm{v}': \bm{v}' \ne \bm{v}} \sum_{(\bm{x}, \bm{y}) \in \mathcal{X}^m \times \mathcal{Y}^m} P_{X^m|V^n}(\bm{x}|\bm{v}) W^m(\bm{y}|\bm{x}) \\
&\sum_{\bm{x}': (\bm{x}', \bm{y}) \in S_n(\bm{v}')} \beta'(F_n, q_n)(\bm{v}', \bm{x}') P_{X^m|V^n}(\bm{x}'|\bm{v}') \\
= &\sum_{\bm{v}': \bm{v}' \ne \bm{v}} \sum_{(\bm{x}, \bm{y}) \in \mathcal{X}^m \times \mathcal{Y}^m} P_{X^mY^m|V^n}(\bm{x}, \bm{y}|\bm{v}) \\
&\sum_{\bm{x}': (\bm{x}', \bm{y}) \in S_n(\bm{v}')} \beta'(F_n, q_n)(\bm{v}', \bm{x}') P_{X^m|V^n}(\bm{x}'|\bm{v}') \\
\le &\sum_{\bm{v}' \in \mathcal{V}^n} \sum_{(\bm{x}', \bm{y}) \in S_n(\bm{v}')} \beta'(F_n, q_n)(\bm{v}', \bm{x}') P_{Y^m|V^n}(\bm{y}|\bm{v}) \\
&P_{X^m|V^n}(\bm{x}'|\bm{v}'),
\end{IEEEeqnarray*}
where (a) follows from \eqref{eq:C1ofLemmaFeinsteinBLC} and \eqref{eq:C2ofLemmaFeinsteinBLC}. Therefore, it follows that
\begin{IEEEeqnarray}{Cl}
&\sum_{\bm{v} \in \mathcal{V}^n} P_{V^n}(\bm{v}) B_n(\bm{v}) \IEEEnonumber \\
\le &\sum_{\bm{v}' \in \mathcal{V}^n} \sum_{(\bm{x}', \bm{y}) \in S_n(\bm{v}')} \beta'(F_n, q_n)(\bm{v}', \bm{x}') P_{X^m|V^n}(\bm{x}'|\bm{v}') \IEEEnonumber \\
&\sum_{\bm{v} \in \mathcal{V}^n} P_{V^n}(\bm{v}) P_{Y^m|V^n}(\bm{y}|\bm{v}) \IEEEnonumber \\
= &\sum_{(\bm{v}', \bm{x}', \bm{y}) \in S_n} \beta'(F_n, q_n)(\bm{v}', \bm{x}') P_{Y^m}(\bm{y}) P_{X^m|V^n}(\bm{x}'|\bm{v}') \IEEEnonumber \\
\levar{(a)} &e^{-n\gamma} \sum_{(\bm{v}', \bm{x}', \bm{y}) \in S_n} P_{V^n}(\bm{v'}) W^m(\bm{y}|\bm{x}') P_{X^m|V^n}(\bm{x}'|\bm{v}') \IEEEnonumber \\
\le &e^{-n\gamma},\label{eq:UBofAverageBn}
\end{IEEEeqnarray}
where (a) follows from the inequality
$$
\beta'(F_n, q_n)(\bm{v}', \bm{x}') P_{Y^m}(\bm{y}) \le e^{-n\gamma} P_{V^n}(\bm{v}') W^m(\bm{y}|\bm{x}')
$$
implied by \eqref{eq:GoodSetForDecoding}. Hence from \eqref{eq:AverageErrorFormula}, \eqref{eq:UBofEpsilonV}, \eqref{eq:UBofAverageAn} and \eqref{eq:UBofAverageBn}, it follows that
\begin{IEEEeqnarray*}{rCl}
\epsilon_n &= &\sum_{\bm{v} \in \mathcal{V}^n} P_{V^n}(\bm{v}) \epsilon_n(\bm{v}) \\
&\le &\sum_{\bm{v} \in \mathcal{V}^n} P_{V^n}(\bm{v}) A_n(\bm{v}) + \sum_{\bm{v} \in \mathcal{V}^n} P_{V^n}(\bm{v}) B_n(\bm{v}) \\
&\le &\Pr\{(V^n, X^m, Y^m) \not \in S_n\} + e^{-n\gamma}.
\end{IEEEeqnarray*}
This completes the proof.
\end{proofof}

By Lemma \ref{le:FeinsteinBLC}, we can immediately obtain the main result as follows.

\begin{theorem}\label{th:ReDirectTheoremBLC}
Let $\bm{V} = \{V^n\}_{n=1}^\infty$ be a source and $\bm{W} = \{W^n\}_{n=1}^\infty$ a channel. If for a random linear code $F_n$, a quantization $q_n: \mathcal{V}^n \times \mathcal{U}^l \to \mathcal{X}^m$ and two sequences $\{m_n\}$, $\{\gamma_n\}$ satisfying
$$
\limsup_{n \to \infty} \frac{n}{m_n} \le R
$$
and
$$
\gamma_n > 0, \gamma_n \to 0, \mbox{ and } n\gamma_n \to \infty, \quad \mbox{as } n \to \infty
$$
it holds that
\begin{IEEEeqnarray}{l}
\limsup_{n \to \infty} \Pr\biggl\{\frac{1}{n} \ln \frac{W^m(Y^m|X^m)}{P_{Y^m}(Y^m)} \le \frac{1}{n} \ln \frac{1}{P_{V^n}(V^n)} \IEEEnonumber \\
\qquad +\: \frac{1}{n} \ln \beta'(F_n, q_n)(V^n, X^n) + \gamma_n \biggr\} \le \epsilon,
\end{IEEEeqnarray}
where $X^m$ denotes the channel input generated randomly subject to the conditional probability distribution $P_{X^m|V^n}$ defined by \eqref{eq:RequirementOfQ}, then the system (Figure \ref{fig:Scheme1}) based on $F_n$ and $q_n$ is $(R, \epsilon)$ transmissible, that is, $\limsup_{n \to \infty} \epsilon_n \le \epsilon$.
\end{theorem}

Theorem \ref{th:ReDirectTheoremBLC} is an easy consequence of Lemma \ref{le:FeinsteinBLC}, and hence the proof is omitted here. If $\mathcal{V} = \mathcal{U}$ is a Galois field, it follows from Proposition \ref{pr:GoodLinearCodes} that for each $n$ there exists at least a good linear code $f_n: \mathcal{V}^n \to \mathcal{U}^l$ satisfying
\begin{equation}\label{eq:ConditionOfGLC}
\max_{P \in \mathcal{P}_n(\mathcal{V}) \backslash \{P_{0^n}\} \atop Q \in \mathcal{P}_l(\mathcal{U})\}} \alpha(f_n)(P, Q) \le 1.
\end{equation}
Then by the definition \eqref{eq:DefinitionOfBeta2} in Lemma \ref{le:IndependenceLC2}, we have
$$
\frac{1}{n} \ln \beta'(f_n, q_n)(\bm{v}, \bm{x}) \le 0
$$
for any $q_n: \mathcal{U}^l \to \mathcal{X}^m$, any $\bm{v} \in \mathcal{V}^n$ and any $\bm{x} \in \mathcal{X}^m$. Furthermore, for any conditional probability distribution $P_{X^m|V^n}$ we may easily construct a quantization $q_n$ to simulate it by \eqref{eq:RequirementOfQ}, so the system (Figure \ref{fig:Scheme1}) based on $f_n$ and $q_n$ can achieve the same performance as that indicated by the Direct theorem in \cite[Theorem 3.8.1]{JSCC:Han200300}. In other words, \emph{there exist good linear codes for constructing limit-approaching joint source-channel coding scheme for general sources and channels}, and to be a \emph{good linear code}, the joint spectrum of $f_n$ should satisfy \eqref{eq:ConditionOfGLC}. The condition \eqref{eq:ConditionOfGLC} may be too strict in practice, so we define an \emph{asymptotically good linear code} by
\vspace{-0.03in}
\begin{equation}\label{eq:ConditionOfAGLC}
\limsup_{n \to \infty} \frac{1}{n} \ln \max_{P \in \mathcal{P}_n(\mathcal{V}) \backslash \{P_{0^n}\} \atop Q \in \mathcal{P}_l(\mathcal{U})\}} \alpha(f_n)(P, Q) \le 0.
\end{equation}
\vspace{-0.07in}

\itwsection{Conclusions and Discussion}

A limit-approaching lossless joint source-channel coding scheme based on linear codes is proposed in this paper, and conditions of good linear codes or asymptotically good linear codes are given. The most surprising conclusion of this work is that \emph{the performance of a lossless joint source-channel coding scheme based on linear codes is determined by the codes' joint spectrum instead of the image spectrum}.

The result of this paper is only a part of our ongoing research of joint source-channel coding, and lots of interesting results can be obtained by using the methods and results in this paper. For example, note that the quantization in our scheme is arbitrary, we can easily design a variable rate coding scheme for memoryless channels by constructing an appropriate quantization $q_n$. Besides, the method established in this paper can be extended to the case of multiple access channels with correlated sources, which includes Slepian-Wolf coding as its special case. The details of the whole work is presented in \cite{JSCC:Yang200600}.

\end{itwpaper}


\begin{itwreferences}

\bibitem{JSCC:Cover198011}
T.~M. Cover, A.~E. Gamal, and M.~Salehi, ``Multiple access channels with
  arbitrarily correlated sources,'' \emph{{IEEE} Trans. Inform. Theory},
  vol.~26, no.~6, pp. 648--657, Nov. 1980.

\bibitem{JSCC:Zhong200604}
Y.~Zhong, F.~Alajaji, and L.~L. Campbell, ``On the joint source-channel coding
  error exponent for discrete memoryless systems,'' \emph{{IEEE} Trans. Inform.
  Theory}, vol.~52, no.~4, pp. 1450--1468, Apr. 2006.

\bibitem{JSCC:Csiszar198100}
I.~Csisz\'{a}r and J.~K{\"{o}}rner, \emph{Information Theory: Coding Theorems
  for Discrete Memoryless Systems}.\hskip 1em plus 0.5em minus 0.4em\relax New
  York: Academic Press, 1981.

\bibitem{JSCC:Bennatan200403}
A.~Bennatan and D.~Burshtein, ``On the application of {LDPC} codes to arbitrary
  discrete-memoryless channels,'' \emph{{IEEE} Trans. Inform. Theory}, vol.~50,
  no.~3, pp. 417--437, Mar. 2004.

\bibitem{JSCC:Divsalar199809}
D.~Divsalar, H.~Jin, and R.~J. McEliece, ``Coding theorems for ``{Turbo}-like''
  codes,'' in \emph{36th Allerton Conf. on Communication, Control, and
  Computing}, Sept. 1998, pp. 201--210.

\bibitem{JSCC:Han200300}
T.~S. Han, \emph{Information-Spectrum Methods in Information Theory}.\hskip 1em
  plus 0.5em minus 0.4em\relax Berlin: Springer, 2003.

\bibitem{JSCC:Yang200503}
S.~Yang and P.~Qiu, ``On the performance of linear {Slepian-Wolf} codes for
  correlated stationary memoryless sources,'' in \emph{Proc. DCC 2005},
  Snowbird, UT, Mar. 2005, pp. 53--62.

\bibitem{JSCC:Muramatsu200510}
J.~Muramatsu, T.~Uyematsu, and T.~Wadayama, ``Low-density parity-check matrices
  for coding of correlated sources,'' \emph{{IEEE} Trans. Inform. Theory},
  vol.~51, no.~10, pp. 3645--3654, Oct. 2005.

\bibitem{JSCC:Yang200600}
S.~Yang and P.~Qiu, ``On the performance of lossless joint source-channel
  coding based on linear codes for multiple-access channels,'' in preparation, to be submitted to IEEE
  Trans. Inform. Theory.

\end{itwreferences}

\end{document}